\documentclass[%
 preprint,
 amsmath,amssymb,
 aps,
prd,
]{revtex4-1}
\usepackage[dvipsnames,usenames]{xcolor}
\usepackage{graphicx}
\usepackage{dcolumn}
\usepackage{bm}
\usepackage{mathrsfs,natbib}
\usepackage{graphicx,epsf,amssymb,amsbsy,amsfonts,amssymb,amsmath}
\usepackage{slashed}
\usepackage{comment}
\usepackage{subcaption}
\usepackage{hyperref}

\newcommand{\eq}{\begin{equation}}
\newcommand{\eqe}{\end{equation}}

\newcommand{\beq}{\begin{eqnarray}}
\newcommand{\eeq}{\end{eqnarray}}
\newcommand{\nn}{\nonumber}

\begin{document}

\title{Plasma Effects on Lasing of Uniform Ultralight Axion Condensate}

\preprint{INT-PUB-18-020}

\author{Srimoyee Sen}
\email{srimoyee08@gmail.com}
\affiliation{Institute for Nuclear Theory, University of Washington, Seattle, WA 98105 USA}
\begin{abstract}
Lasing of ultralight axion condensate into photons can be sensitive to the presence of a background plasma owing to its coupling to electromagnetism. Such a scenario is particularly relevant for superradiant axion condensate around stellar mass black holes since the axion mass can be
within a few orders of magnitude of the plasma frequency of the surrounding medium. In this paper I discuss the properties of the plasma around a black hole and analyze its effects on the lasing of a uniform axion condensate of mass of the order of the plasma frequency.

\end{abstract}
\maketitle

\section{Introduction} 
The goal of this paper is to address the question of lasing of an ultralight axion condensate into photons in the presence of a background plasma where the plasma frequency is comparable to the axion mass. The motivation for this analysis is tied to the 
possibility of formation of superradiant axion condensates around spinning black holes which are typically surrounded by the interstellar plasma or accreting material. The mechanism of formation of these condensates is based on the interplay of the Penrose process \cite{Penrose:1969pc} and superradiant instability \cite{1972JETP...35.1085Z, 1971JETPL..14..180Z} where a wave scattering off of a spinning black hole gains in amplitude in the scattering event \cite{PhysRevLett.29.1114, Teukolsky:1974yv}. Such a wave if confined around the black hole can result in exponential growth of the field-amplitude forming a condensate \cite{Press:1972zz}. Due to the Pauli exclusion principle however, fermionic fields are not superradiant \cite{PhysRevLett.31.1265, Chandrasekhar:1976ap, Iyer:1978du} and it is only for massive bosons, axions being one of them, that a Bose-Einstein condensate with a large occupation number can form thereby causing the black hole to spin down\cite{Zouros:1979iw, Detweiler:1980uk, Gaina:1988nf}. For a comprehensive review see \cite{Brito:2015oca}. This phenomenon has attracted significant interest of late in anticipation of upcoming
gravitational wave observations \cite{Brito:2015oca, Brito:2017zvb, East:2017ovw, Arvanitaki:2009fg, Arvanitaki:2010sy, Arvanitaki:2014wva, Baryakhtar:2017ngi, Arvanitaki:2016qwi, Rosa:2016bli, Rosa:2017ury, Conlon:2017hhi} in particular involving stellar mass black holes. For black holes of mass $M$ the typical mass of axions that can condense are of the order of $\frac{1}{GM} \sim 10^{-10} \text{eV}\frac{M_{\odot}}{M}$ where $G$ is the Newton's constant. This allows for the attractive possibility of detection of axions of very low mass through gravitational waves alone \cite{Brito:2017zvb}. There have been analysis of indirect signatures of the superradiant axion condensates utilizing the spin distribution of black holes \cite{Brito:2017zvb, Arvanitaki:2016qwi}, direct detection of gravitational waves through level transitions \cite{Arvanitaki:2016qwi}, finite size effects \cite{Baumann:2018vus} and annihilation to gravitons \cite{Arvanitaki:2016qwi}. However, the prospects of electromagnetic signatures through lasing have not been explored in great detail except in \cite{Rosa:2017ury}. The absence of such analysis for black holes of stellar mass or larger can partly be attributed to the extremely long wavelengths of photons under consideration which are ostensibly undetectable on earth based telescopes. \footnote{ The interstellar medium damps electromagnetic waves of such long wavelengths which have frequencies smaller than the scale of inverse plasma frequency $\sim 10^{-10} - 10^{-12}$ eV}. However, it is important to note that the detectability of gravitational signatures from the axion condensate may critically depend on the non-occurrence of lasing. A particularly gloomy prospect can involve complete depletion of the axion condensate through lasing thereby eliminating any detectable gravitational wave signatures while the photons produced in the process are not observable either. 



One of the key features of this problem is the contrast between the time scales associated with spontaneous and stimulated emissions. 
The spontaneous emission rate for axions, given by $\sim\frac{m_a^3}{f_a^2}$ where $m_a$ and $f_a$ are the axion mass and the axion decay constant, is miniscule for a superradiant axion owing to its small mass. For QCD axions, this rate is further suppressed since $m_a$ and $f_a$ are related by $m_a f_a \sim \lambda_{\text{QCD}}^2$. The associated decay time scale given by $10^{64}\left(\frac{M}{10M_{\odot}}\right)^5 \text{years}$, is much larger than the age of the universe. This may lead one to erroneously declare that the decay of superradiant axions to photons is irrelevant for any time scales of interest. The process of lasing however is largely dictated by the rate of stimulated emission as opposed to the rate of spontaneous emission. As I review in the main text of this paper, in the absence of any matter coupling the rate of stimulated emission is given by $\sim\frac{\beta |\phi|}{\pi f_a} m_a$ for a uniform axion condensate $\phi$ where $\beta/f_a$ is the coupling between photon and axion. The corresponding depletion time scale $\frac{1}{\frac{\beta |\phi|}{\pi f_a}}\frac{M}{10 M_{\odot}}10^{-4}$ seconds, can be much smaller than the time scale associated with spontaneous emission.


It is however important to note that the process of lasing can be sensitive to any non-uniformity in the axion condensate as well as the presence of a plasma around the black hole. The former can obstruct or weaken laser growth due to ``leakage" of photons from the finite-sized condensate \cite{Hertzberg:2018zte}. I postpone the discussion of non-uniformity of superradiant condensates for future work. A similar blocking of the laser can take place due to the frequency and wavelength dependent scattering of photons in matter. This can lead to a kinematic blocking of the two-photon decay mode when plasma mass of the photon is much larger than the axion mass \cite{Preskill:1982cy}. Medium dependence of lasing is particularly subtle when one or more physical scales of the surrounding plasma is comparable to the axion mass scale. The discussion in this paper pays particularly close attention to this regime of the parameter space. As we will see in the subsequent sections, for superradiant axions around stellar mass black holes, the axion mass may indeed be comparable to the plasma frequency of the medium around it. Therefore it is important to analyze this problem in the context of a uniform axion condensate without obscuring the physics of interest with spatial non-uniformities.

Typically lasing of axions to photons is dealt with by solving Maxwell's equations in the presence of a source term driven by the axions. These equations supplemented  with an appropriate term for the conductivity can then be used to describe medium response. In this paper I estimate the conductivity of the medium around black hole and solve the corresponding Maxwell's equations in the presence of a uniform axion condensate. It is evident that the detailed electromagnetic response of interest to this paper is a very involved problem in its full generality which requires numerical work. However, such a treatment is beyond the scope of this paper. Here I work with a simple toy model that can clearly explain the physics in question. I begin with a short review of the superradiant instability and lasing of axions in the absence of any matter. This is followed by estimates of the conductivity around the black hole and solution to the lasing problem in the presence of a finite conductivity. 
\section{Maxwell's equations for the axion cloud}

Before delving into the relevant equations of motions for lasing of axions, let us first review the basics of the superradiant instability. In the process we will also estimate the order of magnitude of some of the parameters involved in the problem. As we will see, the most generic solution to the Maxwell's equations are not very enlightening and one needs to identify small parameters in the problem so as to understand better the conditions under which lasing of axions to photons can take place. As stated earlier, an axion condensate forms around a spinning black hole of mass $M$ when axion mass $m_a$ is of the order of $\frac{m_P^2}{M}$ where $m_P$ is the Planck mass. The condensate is well described by hydrogen atom wave functions with a coupling constant $\alpha_M \sim \frac{m_a M}{m_P^2} $. The corresponding spectrum is given by $\omega\approx m_a (1-\frac{\alpha_M}{2n^2})$ where $n, l, m$ are the hydrogen quantum numbers. The axion cloud extracts angular momentum from a maximally rotating black hole as long as the superradiance condition is satisfied
\beq
\frac{\omega}{m} < \frac{1}{2}r_g^{-1}
\eeq
where $r_g \sim M/m_P^2$. Parametrically the maximal occupation number of a level is given by
$\sim\frac{M^2}{m_P^2}$ \cite{Arvanitaki:2014wva} which coupled with the axion density of $\sim m_a^2 \phi$ and a parametric estimate for the volume of the condensate $\sim \frac{1}{m_a^3}$ can lead to $\frac{|\phi|}{f_a}\approx 1$.

As we will see below the product of the axion-photon coupling $\beta/f_a$ and the axion vev $\phi$ given by $\sim \frac{\beta \phi}{f_a}$, is one of the possible small parameters which we will eventually expand in to make sense of the results.  
 With this brief review let us now
write down the axion-photon Lagrangian \cite{Sikivie:1983ip, Arvanitaki:2009fg}
\beq
\mathcal{L}&=&-\frac{1}{4}F_{\mu\nu}F^{\mu\nu}+\frac{C \beta}{4\pi f_a}\phi \epsilon^{\mu\nu\lambda\rho}F_{\mu\nu}F_{\lambda\rho}\nn\\
&& \,\, \frac{1}{2}\partial_{\mu}\phi\partial^{\mu}\phi - \frac{1}{2}m_a^2 \phi^2 .\nn
\eeq
Here $C$ is some model dependent numerical constant \cite{Arvanitaki:2009fg}. 
Ignoring the back-reaction of the electromagnetic field on the evolution of axion field and assuming a spatially uniform axion condensate the
equations of motion can be expressed as 
\beq
\nabla \times \mathbf{B} -\frac{d \mathbf{E}}{dt}=-\frac{C \beta}{\pi f_a}\frac{d\phi}{dt}\mathbf{B}.
\label{maxnorm}
\eeq
In order to describe the physics of lasing one has to express the electric, magnetic and axion fields in their second quantized form in terms of creation and annihilation operators and then solve for their expectations values in coherent states according to Eq. \ref{maxnorm}. It is the expectation values of the gauge fields in coherent states that are expected to exhibit exponential growth as a signature of lasing. Let us now write down the second quantized gauge field  
\beq
\mathbf{A}(\mathbf{r},t)=\frac{1}{2\sqrt{V}}\sum_\mathbf{k}\left(\hat{\boldsymbol{\alpha}}_{\mathbf{k}}(t)e^{i (\mathbf{k}.\mathbf{r}-\omega_k t)}+\hat{\boldsymbol{\alpha}}^*_{\mathbf{k}}(t)e^{-i (\mathbf{k}.\mathbf{r}-\omega_k t)}\right)
\label{2ndq}
\eeq
as well as a spatially uniform second quantized axion field 
\beq
\phi(\mathbf{r},t)=\left(\hat{\phi} e^{-i m_a t}+\hat{\phi}^*e^{i m_a t}\right).
\label{2ndq2}
\eeq
Here, the time dependence of $\hat{\boldsymbol{\alpha}}_{\mathbf{k}}(t)$ is slow compared to the photon frequency of $\omega_k$. This time dependence is intended to eventually capture the physics of exponentially growing laser. On the other hand, ignoring the back-reaction of the gauge field on the axion condensate leads to $\hat{\phi}$ being independent of time. 
Let us now concentrate on a particular mode of the gauge field given by 
\beq
\hat{\boldsymbol{\alpha}}_\mathbf{k}(t)=\delta_{\hat{\mathbf{k}},\hat{z}}\delta_{\omega_k,\frac{m_a}{2}} \hat{\alpha}^x_k(t) \hat{x}+
\delta_{\hat{\mathbf{k}},-\hat{z}}\delta_{\omega_k,\frac{m_a}{2}} \hat{\alpha}^y_k(t) \hat{y}.
\label{2ndq3}
\eeq
Substituting Eq. \ref{2ndq2} and \ref{2ndq3} in Eq. \ref{maxnorm}, I find
\beq
k^2 \langle\boldsymbol{\alpha}^{x/y}_{k}(t)\rangle + \langle\ddot{\boldsymbol{\alpha}}^{x/y}_{k}(t)\rangle-i m_a\langle\dot{\boldsymbol{\alpha}}^{x/y}_{k}(t)\rangle-\frac{m_a^2}{4}(t)\langle \boldsymbol{\alpha}^{x/y}_{k}(t)\rangle = -\frac{C \beta}{\pi f_a}k m_a\langle \phi (\boldsymbol{\alpha}_k^{y/x}(t))^*\rangle .
\label{finmax2}
\eeq
  Eq \ref{finmax2} which excludes medium response will produce lasing solutions with the following ansatz 	
	\beq
	\langle\alpha_k^x(t)\rangle=f(t), \langle\alpha_k^y(t)\rangle=f(t)e^{i\theta}, \langle\phi\rangle=\phi_0^R
	\label{ansatz1}
	\eeq
	where $\phi_0^R$ is a real constant and $f(t)$ is a complex valued spatially uniform function of time given by
	\beq
	f(t)=f_0(1+i \kappa)e^{\lambda t}
	\label{ansatz2}
	\eeq
	with $f_0$, $\kappa$ and $\lambda$ being real numbers.
	It is easy to see that the phase shift between 
	the two orthogonal components of the gauge field can be adjusted to absorb any $U(1)$ phase of the axion field operator expectation in the   coherent state. This is why I choose $\langle\phi\rangle$ to be real without any loss of generality. 
	
	Eq \ref{finmax2} can be solved for any wavelength $k$ which will relate the growth rate $\lambda$ to the wave number $k$, the axion mass and the axion condensate field $\phi_0^R$.	The growth rate is found to be maximum at $k=\frac{m_a}{2}$ for $\frac{C \beta |\phi|}{\pi f_a}\ll 1$.
	Solving Eq. \ref{finmax2} for	$|k|=\omega=m_a/2$, the growth rate can be written as
	$\lambda \approx m_a\frac{C \beta}{2\pi f_a}|\phi_0^R|$ in the limit of $\frac{C \beta |\phi|}{\pi f_a} \ll 1$. For $\theta=0, \pi$, in the limit of $\frac{C \beta |\phi|}{\pi f_a} \ll 1$, one finds $|\kappa|\approx 1$. Similarly for $\theta=\pm \frac{\pi}{2}$, the growing modes correspond to $|\kappa|\sim \frac{4\pi f_a}{C \beta |\phi|}$ and  $|\kappa|\sim \frac{C \beta |\phi|}{4\pi f_a}$. 
	
	\section{Estimates of the Plasma Scales}
	The solution in the previous section of course ignores medium effects completely. Incorporating these effects in the equations of motion in principle involves augmenting the equations with appropriate constitutive relations for the current density $j_{\text{medium}}$ in linear response to an external electromagnetic field. Such a linear response current is well approximated by $j_{\text{medium}}=\sigma \mathbf{E}$ where $\sigma$ is the electrical conductivity of the medium. The response to electromagnetic waves typically carries frequency dependence unless one is in the collision dominated regime. To be specific, the dependence of the electrical conductivity on the frequency of interest can be approximately expressed as
	\beq
	\sigma(\omega)=\frac{4\pi n_e e^2 \tau_{\text{coll}}}{m_e(1-i\tau_{\text{coll}} \omega)}
	\eeq
	where $n_e$ is the density of electrons in the medium, $m_e$ is the mass of an electron and $\tau_{\text{coll}}$ is the inverse collision frequency. 
	In the limit $\omega\tau_{\text{coll}} \ll 1$, one is in the collision dominated regime where conductivity is independent of the frequency and is given by $\frac{ 4\pi n_e e^2 \tau_{\text{coll}}}{m_e}$. In the collision-less limit with $\omega\tau_{\text{coll}} \gg 1$, the conductivity is frequency dependent and is given by $\frac{i 4\pi n_e e^2 }{\omega m_e}$. 
	
	In principle one can solve Maxwell's equations as a function of an arbitrary conductivity. However, this is cumbersome and unnecessary in the present context. Instead it is much more useful to begin with an estimate of the conductivity in regions around the black hole which simplifies the equations significantly. With this in mind, I now proceed to estimate the conductivity of the black hole environment. In the presence of a thin accretion disc, the region around a black hole outside of the disc is expected to resemble the interstellar medium to first approximation. Interstellar medium typically consists of hot ionized hydrogen plasma with a density of about $1/\text{cc} - 0.001/\text{cc}$ and temperature of about $10^4 \text{K} - 10^6 \text{K}$. In order to determine the conductivity of this region I have to first estimate the collision frequency of electrons in it. The collision frequency is related to the mean free path of electrons which is given by 
\beq
\lambda_{\text{mfp}} &=& \frac{T^2}{n_e \pi e^4 \ln(\Lambda)}
\label{mfp}
\eeq 
for coulomb collisions.
Here $T$ is the temperature, $n_e$ is the density of electrons and the Coulomb logarithm $\ln(\Lambda)\sim 10$. Assuming the density to be one particle per cubic centimeter $n_e\sim 8 \times 10^{-15} \text{eV}^3$and a temperature of $10^4 \text{K} \sim 1 \text{eV}$, the mean free path is given by $\lambda \sim 4.72 \times 10^{14} \left(\frac{1/\text{cc}}{n_e}\frac{T^2}{(1 \text{eV}^2)}\right)\text{eV}^{-1}$. The collision frequency is related to the mean free path as $\nu_{\text{coll}} \sim v_e(\lambda)^{-1}$ where $v_e$ is the
speed of an electron $v_e \sim \sqrt{T/m_e}$. For the temperature and densities under consideration $v_e \sim 10^{-3}$ and the collision frequency is $\sim 10^{-18}\left(\frac{1 \text{eV}}{T}\right)^{3/2}\frac{n_e}{1/\text{cc}} \text{eV}$. Since the frequency of the lasing photons is set by the axion mass $\sim 10^{-11}\text{eV}\frac{10 M_{\odot}}{M}$, it is clear that one is in the collision-less limit outside of the accreting disc. The conductivity is then given by $i \frac{4 \pi n_e e^2}{\omega m_e}$ upto corrections of the order $m_a \tau_{\text{coll}}$. Maxwell's equations in the collisionless limit with $\phi=0$ has propagating modes only for frequencies larger than the plasma frequency given by $\omega_P^2=\omega \text{Im}[\sigma] = \frac{4\pi n_e e^2}{m_e}$. The corresponding dispersion relation is given by $\omega^2=k^2+\omega_P^2$. 

Similarly an estimate of the collision frequency can be obtained if there is accreting matter around a black hole. In the Eddington limit \cite{Coppi} the quantity of matter around a black hole is related to its mass $M$, the Schwarzschild radius $r_g$ and accretion time (also known as Salpeter time \cite{Coppi}) $\tau_{\text{accr}}$ as $\delta M \sim M \frac{r_g}{\tau_{\text{accr}}}$ \cite{Arvanitaki:2009fg}. The accretion time is given by $\tau_{\text{accr}}=\frac{\sigma_T}{4\pi G m_{\text{proton}}}$ where $\sigma_T$ is the Thomson cross section and $m_{\text{proton}}$ is the proton mass. The accretion time evaluates to about $4\times 10^8 \text{years}=10^{40} \text{GeV}^{-1}$ which can then be used to estimate the quantity of matter around the black hole 
\beq
\delta M \sim 4\times 10^{-22}\left(\frac{M}{M_{\odot}}\right).
\eeq
The accreting matter can be assumed to be ionized hydrogen to a good approximation. To determine the mean free path of electrons in the accreting matter, one needs to know the density of electrons in it as well as its temperature. It is known that in the Eddington limit, assuming black body radiation, the temperature of the accreting matter around a black hole at a radial distance $r$ is given by $T=\left(\frac{3 M^2 G_N}{8\pi r^3 \tau_{\text{accr}}}\right)^{1/4}$ which estimates to about $\sim 500 \left(\frac{10 M_{\odot}}{M}\right)^{1/4}\text{eV}$ at the ISCO ($r\sim 6 G_N M$), for a black hole of mass $10 M_{\odot}$. Keeping in mind that most of the accreting matter is hydrogen one can make a rough estimate of the density of electrons as 
\beq
n_e &\sim & \frac{\delta M}{m_{\text{proton}}\frac{4}{3}\pi r^3}\nn\\
&\sim& \frac{4\times 10^{-20}\left(\frac{M}{10 M_{\odot}}\right)^2 M_{\odot}}{m_{\text{proton}}\frac{4}{3}\pi r^3}\nn\\
&\approx& 4.42\times 10^{-17}\left(\frac{10 M_{\odot}}{M}\right)^{-1}\text{MeV}^3.
\eeq
The mean free path of electrons in the accreting matter can now be readily obtained using 
\beq
\lambda &=& \frac{T^2}{n_e \pi e^4 \ln(\Lambda)}\nn\\
&\approx & \frac{0.5^2\times 10^{-6}}{4.42\times 10^{-17}\pi (\frac{4\pi}{137})^2 10}\left(\frac{M}{10 M_{\odot}}\right)^{1/2} \text{MeV}^{-1}\nn\\
&\sim & 2.14 \times 10^{10}\left(\frac{M}{10 M_{\odot}}\right)^{1/2}\text{MeV}^{-1}
\eeq
and the corresponding collision frequency is given by
\beq
\omega_{\text{coll}}^{\text{accr}}\sim\frac{\sqrt{T/m_e}}{\lambda}\sim 10^{-6}\text{eV}.
\eeq
Hence we see that in the accreting region, $\omega/\omega_{\text{coll}}^{\text{accr}} \ll 1$ for $\omega \sim m_a \sim 10^{-11}$ \text{eV}
and the conductivity is 
 \beq
\sigma_{\text{accr}} &=& \frac{(4\pi n_e e^2\tau)}{m_e}\nn\\
&=& (4\pi)\frac{T^{3/2}}{\pi e^2  m^{1/2}\log(\Lambda)}\nn\\
&\approx& 70\left(\frac{10 M_{\odot}}{M}\right)^{3/8}\text{eV}.
\label{cond2}
\eeq
In the absence of an axion condensate the collision dominated limit of Maxwell's equations is accompanied by damping of electromagnetic waves. In this regime for a conductivity of $\sigma$, the characteristic length over which electromagnetic waves are damped is given by $\frac{\sigma}{2}$ for $\sigma \ll \omega$ and $\sqrt{\sigma\omega}/\sqrt{2}$ for $\omega\ll\sigma$.

\section{Axion Lasing in Matter Background}
Having obtained an estimate of the conductivity we can now solve Maxwell's equations coupled to an axion condensate in the presence of a finite conductivity 
\beq
\nabla \times \mathbf{B} -\frac{d \mathbf{E}}{dt}=\sigma \mathbf{E}-\frac{C \beta}{\pi f_a}\frac{d\phi}{dt}\mathbf{B}.
\label{max2}
\eeq
One could again use Eq. \ref{2ndq}, Eq. \ref{2ndq2}, Eq. \ref{2ndq3} to write an equation for the expectation values for the creation and annihilation operators just as in Eq. \ref{finmax2}
\beq
k^2 \langle\boldsymbol{\alpha}^{x/y}_{k}(t)\rangle &+& \langle\ddot{\boldsymbol{\alpha}}^{x/y}_{k}(t)\rangle-i m_a\langle\dot{\boldsymbol{\alpha}}^{x/y}_{k}(t)\rangle-\frac{m_a^2}{4}(t)\langle \boldsymbol{\alpha}^{x/y}_{k}(t)\rangle\nn\\
 &=&\sigma \left(\langle\ddot{\boldsymbol{\alpha}}^{x/y}_{k}(t)\rangle-i \frac{m_a}{2}\langle\dot{\boldsymbol{\alpha}}^{x/y}_{k}(t)\rangle\right) -\frac{C \beta}{\pi f_a}k m_a\langle \phi (\boldsymbol{\alpha}_k^{y/x}(t))^*\rangle .
\label{finmax3}
\eeq

The ansatz of Eq. \ref{ansatz1} and \ref{ansatz2} solve Eq. \ref{finmax3} as well. As stated earlier the solutions are not particularly enlightening as a function of an arbitrary conductivity. Relatively simple expressions are obtained in the strictly collision-less and the strictly collision dominated limit. From the estimates of the conductivity above, it is clear that most of the black hole environment for a thin accretion disc is extremely well described by the collision-less limit. Similarly, the accretion disc is very well described by the collision dominated limit. 
Let us first consider Eq. \ref{finmax3} in the collision dominated regime relevant for the accretion disc. It is important to note that in the accretion disc, the conductivity as obtained in Eq. \ref{cond2} is much larger than the axion mass scale. Plugging the ansatz of Eq. \ref{ansatz1} and \ref{ansatz2} in Eq. \ref{finmax3} one can see that $|\sigma\mathbf{E}|$ is parametrically larger (by a factor of $\frac{\sigma_{\text{accr}}}{m_a}$) compared to the rest of the terms. As a result the gauge field solution is strictly zero and no lasing can take place inside a plasma modeling the accreting region. 

Let us now concentrate on the region outside of the accretion disc for a thin disc. As we will see in this section, the possibility of lasing in the collision-less limit depends on the relative magnitude of the plasma frequency to the axion mass and that indeed they can be comparable in a region of the parameter space relevant to superradiant condensates. In this regime, the growth rate of the laser as well as possible wavelengths exhibiting such growth depend on the strength of the axion-photon coupling, the axion condensate and the difference between the plasma mass and the axion mass.

\begin{figure}
    \centering
    \begin{subfigure}[b]{0.4\textwidth}
        \includegraphics[width=\textwidth]{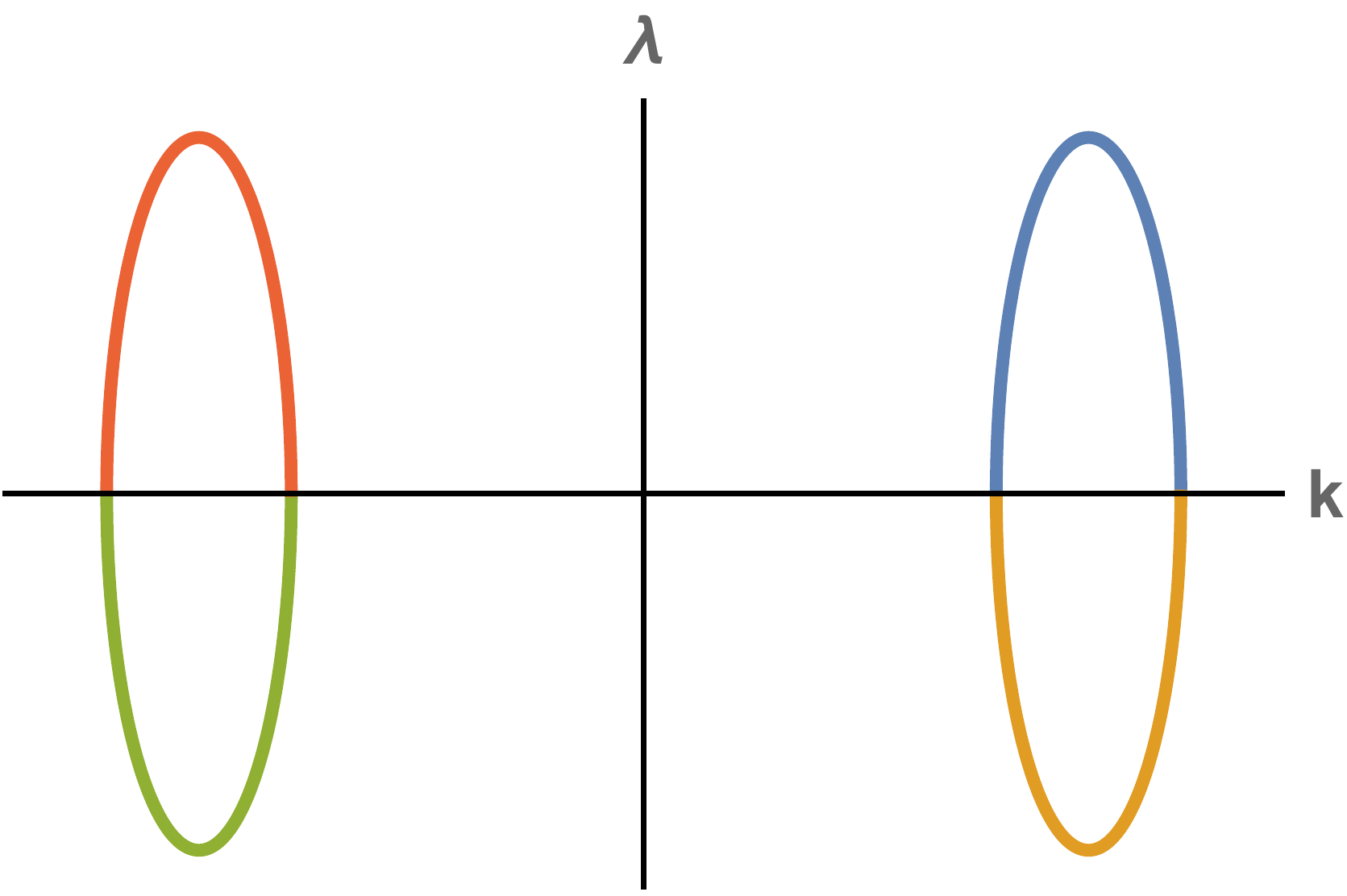}
        \caption{Growth rate vs wave number for $(\xi|\phi|)^2\ll \frac{m_a^2-4\omega_P^2}{m_a^2}$ with $\xi|\phi|< 1$, $(\sqrt{m_a^2-4\omega_P^2}/m_a)<1$}
        \label{fig1}
    \end{subfigure}
    ~ 
    \begin{subfigure}[b]{0.4\textwidth}
        \includegraphics[width=\textwidth]{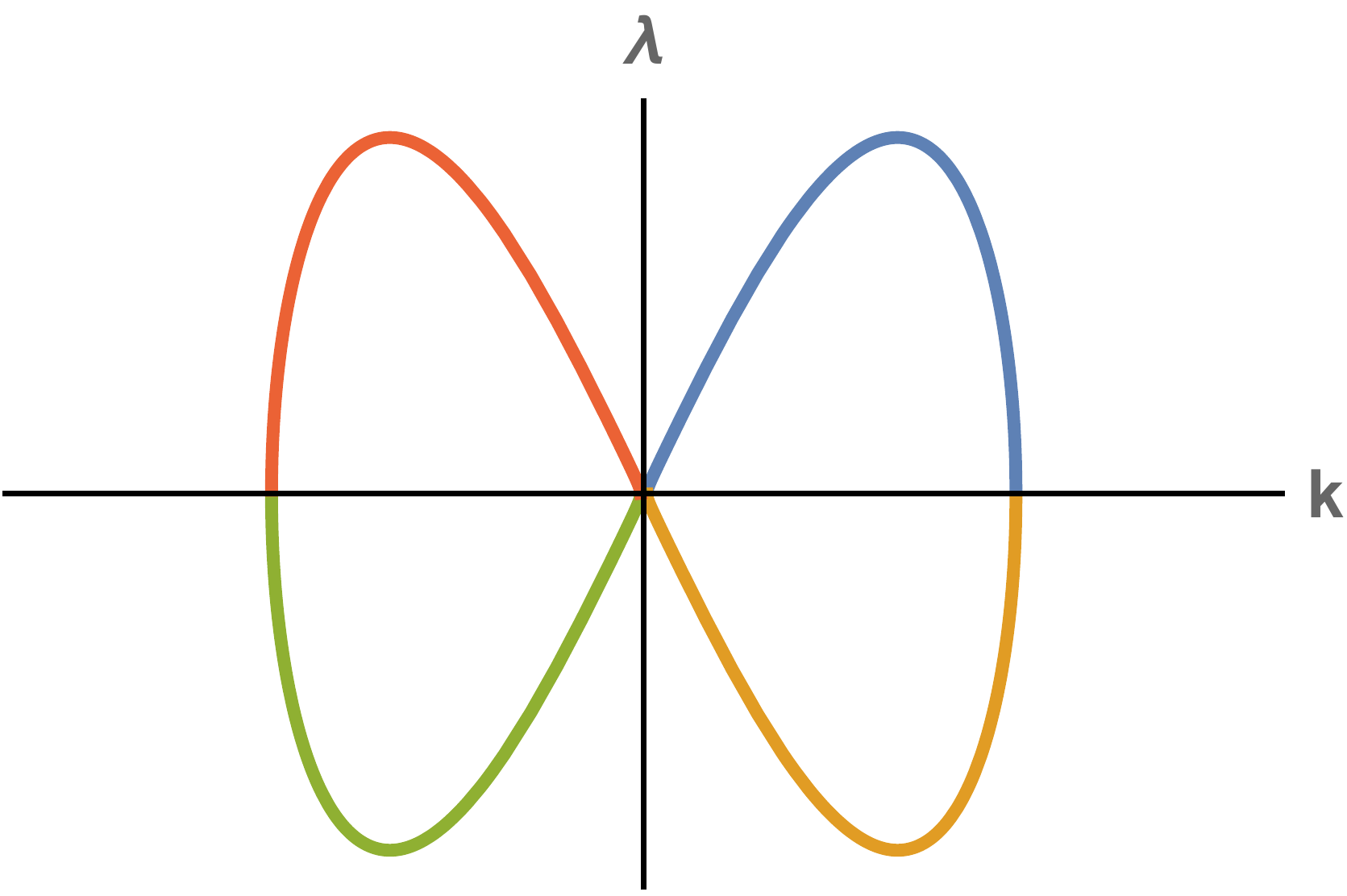}
       \caption{Growth rate vs wave number for $(\xi|\phi|)^2\gg \frac{m_a^2-4\omega_P^2}{m_a^2}$ with $\xi|\phi|< 1$, $(\sqrt{m_a^2-4\omega_P^2}/m_a)<1$}
        \label{fig2}
    \end{subfigure}
    ~ 
    \caption{I plot the solution to the Eq. \ref{max2} and Eq. \ref{finmax3} in two regimes of interest set by $\xi|\phi|< \sqrt{\frac{m_a^2-4\omega_P^2}{m_a^2}}$ and $\xi|\phi|> \sqrt{\frac{m_a^2-4\omega_P^2}{m_a^2}}$ with $\xi|\phi|<1$ and $(\sqrt{m_a^2-4\omega_P^2}/m_a)<1$. The four different colors correspond to different branches of the solution.}\label{fig:12}
\end{figure}

Analyzing Eq. \ref{max2} and Eq. \ref{finmax3} in the collision-less limit leads to exponentially growing solutions as expected. Just like the solution of Eq.\ref{finmax2}, solving Eq. \ref{finmax3} will lead to a relation between the wavelength of the photons, the axion mass, the axion condensate and the lasing rate. Defining $\frac{C\beta}{\pi f_a}\equiv \xi$, I find
exponential growth for momentum  
\beq
\frac{-m_a \xi |\phi|+\sqrt{m_a^2+m_a^2\xi^2|\phi|^2-4\omega_P^2}}{2}<|\mathbf{k}|<\frac{m_a \xi |\phi|+\sqrt{m_a^2+m_a^2\xi^2|\phi|^2-4\omega_P^2}}{2}.
\label{kconst1}
\eeq
 I analyze the laser growth rate in the two limits given by $ \xi |\phi|> \sqrt{m_a^2-4\omega_P^2}/m_a$ and $\sqrt{m_a^2-4\omega_P^2}/m_a > \xi |\phi|$ where both $\xi|\phi|$ and $(\sqrt{m_a^2-4\omega_P^2}/m_a)$ are taken to be less than $1$. For the small $\xi|\phi|$ limit ($\sqrt{m_a^2-4\omega_P^2}/m_a > \xi |\phi|$), the maximum growth rate is found to be at $|\mathbf{k}|\equiv k_{max,1}\approx \sqrt{\frac{m_a^2}{4}-\omega_P^2}$. The corresponding growth rate is given by 
\beq
\lambda_{k_{max,1}}\approx\frac{m_a^2\sqrt{m_a^2-4\omega_P^2}}{2 m_a^2-4\omega_p^2}\left(\xi|\phi|\right)=\frac{m_a^2\sqrt{m_a^2-4\omega_P^2}}{2 m_a^2-4\omega_p^2}\left(\frac{C\beta}{\pi f_a}|\phi|\right).
\label{growth1}
\eeq
 The expansion in $|\phi|\xi$ for the maximum growth rate $\lambda_{k_{max,1}}$ breaks down when $\xi |\phi| \sim \frac{\sqrt{m_a^2-4\omega_P^2}(m_a^2-2\omega_P^2)}{\sqrt{2}m_a\omega_P^2}$. This gives the maximum possible growth rate in the small $|\phi| \xi$ limit to be 
\beq
\lambda^{MAX}_1 \sim \frac{m_a(m_a^2-4\omega_P^2)}{2\sqrt{2}\omega_P^2}.
\label{Max1}
\eeq 
 These growing modes are given by $\{\theta=0, \kappa \approx -1\}$, $\{\theta=\pi, \kappa \approx 1 \}$, $\{ \theta=\frac{\pi}{2}, \kappa\approx \frac{4(m_a^2-2\omega_p^2)^2}{m_a^3 \frac{C \beta \phi}{\pi f_a}\sqrt{m_a^2-4\omega_P^2}}\}$ and $\{\theta=\frac{-\pi}{2}, \kappa\approx -\frac{m_a^3 \frac{C \beta \phi}{\pi f_a}\sqrt{m_a^2-4\omega_P^2}}{4(m_a^2-2\omega_p^2)^2}\}$. Similarly in the limit $\sqrt{\frac{m_a^2-4\omega_P^2}{m_a}}<\xi|\phi|$, maximum growth rate is achieved at
$|\mathbf{k}|\equiv k_{max,2} \approx m_a \xi|\phi|\sqrt{\frac{1+2 \xi^2 |\phi|^2}{2+8 \xi^2|\phi|^2}}$ and the maximum growth rate is given by 
\beq
\lambda_{k_{max,2}}\approx \frac{m_a \xi^2|\phi|^2(\sqrt{2+4\xi^2|\phi|^2}-\sqrt{1+4\xi^2|\phi|^2})}{\sqrt{(1+4\xi^2|\phi|^2)(3+8\xi^2|\phi|^2-2\sqrt{2}\sqrt{1+6\xi^2|\phi|^2+8\xi^4|\phi|^4})}}. 
\label{Max2}
\eeq
One can see that the growth rate given by Eq. \ref{Max2} which is obtained in the limit $\sqrt{\frac{m_a^2-4\omega_P^2}{m_a}}<\xi|\phi|$, is independent of the plasma frequency 
whereas the growth rate in the limit of small $\xi|\phi|$ given by Eq. \ref{growth1} is not.
Fig. \ref{fig:12} is a schematic presenting the growth rate as a function of the wave number in two possible regimes of interest : $\xi|\phi|< \frac{m_a^2-4\omega_P^2}{m_a^2}$ and $\xi|\phi|> \frac{m_a^2-4\omega_P^2}{m_a^2}$. As can be seen from the figure, besides the growing modes, there exist damped modes with damping rates identical to the growth rates of the growing modes. These damped and growing modes correspond to different branches of the solution to Eq. \ref{finmax3}. For the purpose of lasing of course, the damped modes are of no consequence. 

As can be seen from the inequality of \ref{kconst1}, there is no growing mode when $\omega_p > \frac{\sqrt{m_a^2(1+\xi^2|\phi|^2)}}{2}$. For $\xi|\phi|$ of the order of $\sim 1$ in the limit of $\xi |\phi|> \sqrt{\frac{m_a^2-4\omega_P^2}{m_a^2}}$ the maximum growth rate is parametrically given by $\sim m_a$. For fixed $\xi|\phi|$, lasing becomes more and more disfavored as the black hole mass increases which drives the axion mass to smaller values thus forcing $m_a^2(1+\xi^2|\phi|^2)< 4 \omega_P^2$ eventually.  
  Plugging in the density for the ISM I find that the plasma frequency is given by $\omega_p^2 \sim 20 \times 10^{-21} \frac{n_e}{1/\text{cc}} \text{eV}^2$. Comparing the plasma frequency with frequency of the photon $\sim 10^{-11}\frac{10 M_{\odot}}{M} \text{eV}$ one can conclude that lasing of a spatially uniform condensate of superradiant axions around low mass stellar black holes ($10 M_{\odot}$) can take place for background densities smaller than $\leq  10^{-3}/\text{cc}$. Note that the corresponding growth rate of the electromagnetic field is about $\sim 10^{-4}\left(\frac{10^{-11}\text{eV}}{m_a}\right)$ seconds for $\xi|\phi|\sim 1$. 

\section{Outlook}

In this paper I address the question of lasing of a uniform axion condensate in superradiant mass-range in the presence of matter coupling to Maxwell's equations. Outside of the accreting region for stellar mass black holes, densities can be sufficiently small so as to allow for the possibility of lasing. The lasing time scale can be shorter than a second for a condensate $|\phi| \sim f_a$ with order $\sim 1$ coupling to photon. In the case of an extremely small lasing time scale, there is a possibility of depleting the condensate before detection via gravitational waves. Similarly, the depletion time scale can become larger when $|\phi|\beta \ll f_a$ in which case the phenomenon of lasing can leave its imprints in gravitational signatures. 


In order to understand lasing of superradiant condensates in greater detail the idealizations used in this paper need to be relaxed to include realistic modeling of the black hole environment as well as the presence of non-uniformity in the axion condensate. It is expected that the inclusion of non-uniformity in the condensate will increase its depletion time-scale. The time dependence of the strength of the axion condensate as well as the curvature of the metric around a black hole may also play a role in determining the details of lasing process. In principle axion lasing can take place while the superradiant instability itself is arising and an accurate analysis of this will have to involve solving coupled equations describing simultaneous growth or decay of both superradiance and the lasing processes. 

\section{acknowledgement}
I would like to thank Masha Baryakhtar, David B. Kaplan, Robert Lasenby, Sanjay Reddy and Tom Quinn for insightful discussions. This work was supported by U.S. Department of Energy under grant Contract Number DE-FG02-00ER41132. 
\bibliographystyle{apsrev4-1}
\bibliography{axion}
\end{document}